\documentclass[fleqn,usenatbib]{mnras}

\usepackage{newtxtext,newtxmath}

\usepackage[T1]{fontenc}
\usepackage{ae,aecompl}

\usepackage{lipsum}
\usepackage{graphicx}
\usepackage{amsmath}
\usepackage{amssymb}
\usepackage{subdepth}
\usepackage{multicol, multirow}
\usepackage{natbib}
\usepackage{booktabs}
\usepackage{tablefootnote}
\usepackage{threeparttable}
\usepackage[normalem]{ulem}

\newcommand{\pcm}{\,cm$^{-3}$\,}

\newif\ifproofread

\newcommand{\R}[1]{%
\ifproofread
\textcolor{red}{#1}%
\else
#1%
\fi
}

\title[Radio afterglows from BNS mergers and FRBs]{Detectability of radio afterglows from binary neutron star mergers and implications for fast radio bursts}

\author[H. Lin et al.]{Haoxiang Lin,$^{1}$\thanks{E-mail: haoxiang@astron.s.u-tokyo.ac.jp} Tomonori Totani,$^{1,2}$ \\
$^{1}$Department of Astronomy, School of Science, The University of Tokyo, Tokyo 113-0033, Japan\\
$^{2}$Research Center for the Early Universe, School of Science, The University of Tokyo, Tokyo 113-0033, Japan
}

\date{Accepted XXX. Received YYY; in original form ZZZ}

\pubyear{2018}

\begin{document}
\label{firstpage}
\pagerange{\pageref{firstpage}--\pageref{lastpage}}
\maketitle

\proofreadfalse

\begin{abstract}
Binary neutron star (BNS) mergers are one of the proposed origins for both repeating and non-repeating fast radio bursts (FRBs), which associates FRBs with gravitational waves and short gamma-ray bursts (GRBs).  
In this work, we explore detectability of radio afterglows from BNS mergers and compare it to the observed radio limits on FRB afterglow. 
We calculate the afterglow flux powered by the two components: a relativistic jet and a slower isotropic ejecta, and quantify the detection probability as a function of the source redshift, observing time, and flux sensitivity. 
The model parameter distributions inferred from short GRB afterglows are adopted, and viewing angle distributions (uniform spherical, gravitational-wave, on-axis biased) are assumed to reflect different searching scenario. 
Assuming that FRBs are not strongly beamed, we make comparison to FRBs detected with reported radio limits and find the detection probabilities are 1--10\% in general, and hence not a strong constraint on the BNS progenitor model considering the small sample number (<10).
In particular for some nearby FRBs (e.g. 180916.J0158+65, 190608), we find a high chance of detection (>20\% at 10$\mu$Jy sensitivity) for the isotropic component that would peak around $\sim$1--10 years after the merger.  
Therefore a long term radio monitoring of persistent radio emission for these objects is important. 
\end{abstract}


\section{Introduction} \label{sec:intro}
\R{The first detection of gravitational waves (GWs) from binary neutron star (BNS) merger GW170817 was followed by a short gamma-ray burst, an r-process kilonova, and a long-lasting multi-wavelength afterglow powered by the GRB jet \citep[][and references therein]{2017PhRvL.119p1101A, 2017ApJ...848L..12A, 2017ApJ...848L..13A}. 
Continuous monitoring of the afterglow has revealed an angular structure of the relativistic jet \citep[][and references therein]{2019MNRAS.489.1919T, 2019ApJ...886L..17H, 2020arXiv200601150T}. 
On the other hand, slowly-fading radio flares are also expected from the collision of the sub-relativistic kilonova ejecta with the ambient matter \citep{2011Natur.478...82N}, despite no significant detection yet reported.
Such emission, which we will refer to as the {\it ejecta afterglow} analogous to the {\it jet afterglow}, is emitted isotropically and last for years after merger, making it a promising target for follow-up observations \citep{2016ApJ...831..190H, 2019MNRAS.485.4150B}.
}

Fast radio bursts (FRBs) are radio transients with an intrinsic pulse width of milliseconds, whose origin is still enigmatic \citep{2013Sci...341...53T}. 
Nearly 100 FRBs have been discovered \citep{2016PASA...33...45P}\footnote{\url{http://www.frbcat.org}} since the first event archived in 2001 and reported in 2007 \citep{2007Sci...318..777L}, with an all sky rate $\sim$ 10$^{3}$ -- 10$^{4}$ sky$^{-1}$day$^{-1}$ above a few Jy ms \citep[e.g.][and references therein]{2019A&ARv..27....4P}.

Their large dispersion measures (the integrated column density of free electrons along the line of sight) exceeding the contribution by electrons in the Milky Way Galaxy, as well as the direct localization of host galaxies of some FRBs, point to an extragalactic origin and implies source redshift in the range of $z=0.1$--$1$ \citep{2017Natur.541...58C, 2017ApJ...834L...8M, 2017ApJ...834L...7T, 2019Sci...365..565B, 2019Sci...366..231P, 2019Natur.572..352R, 2019ARA&A..57..417C, 2019ApJ...885L..24C, 2019ApJ...883...40L}.
While some of the FRBs are found to be repeating sources \citep{2016Natur.531..202S, 2019Natur.566..235C, 2019ApJ...885L..24C}, most seem to appear only once despite the intensive searches for possible repeating signals \citep{2007Sci...318..777L, 2015MNRAS.454..457P, 2018Natur.562..386S}. 
The distribution of intrinsic pulse time width is different for repeating and non-repeating FRBs, as observed by CHIME, which implies different origins of these two populations \citep{2019ApJ...885L..24C}. \R{The recent discovery of FRB 200428 from the Galactic magnetar SGR 1935+2154 strongly suggests a connection between magnetars and at least some FRBs, although some properties of the radio burst are not the same as extragalactic FRBs \citep{2020arXiv200510324T, 2020arXiv200510828B, 2020ApJ...898L..29M, 2020arXiv200511071L, 2020arXiv200511178R, 2020arXiv200512164T}.}

% FRB & long GRB {2019MNRAS.489.3643M, 2020ApJ...894L..22W}

The high power and short timescale of FRBs naturally associate them with compact stars \citep[see reviews by][]{2019ARA&A..57..417C, 2019PhR...821....1P}, especially neutron stars because of their enormous gravitational and electromagnetic energies and short characteristic time scales of $\mathcal{O}(GM/c^3) \sim 10\mu$s.  
One scenario is that a non-repeating FRB is produced at the time of BNS merger \citep{2013PASJ...65L..12T, 2014ApJ...780L..21Z, 2016ApJ...822L...7W, 2018PASJ...70...39Y}, while repeating FRBs may also be powered by the surviving remnant from the merger \citep{2018PASJ...70...39Y, 2019ApJ...886..110M}. 
\R{Recent localization of a few non-repeating FRBs to massive, modest star-forming host galaxies \citep{2019Sci...366..231P, 2019Natur.572..352R, 2019Sci...365..565B, 2020ApJ...895L..37B, 2020Natur.581..391M, 2020arXiv200513157S, 2020arXiv200513158C} gives a support to the BNS origin of FRBs. }

\R{The currently estimated BNS merger rate at $z=0$ is 110--3840 Gpc$^{-3}$yr$^{-1}$ (90\% confidence) by LIGO/VIRGO \citep{2019PhRvX...9c1040A}, which can be translated to an all-sky rate of $\sim$60--2100 sky$^{-1}$day$^{-1}$ if integrated to $z=1$ with the cosmic star formation rate evolution convolved with the delay time distribution of BNS mergers \citep[e.g.][]{2018PASJ...70...39Y}. Some papers have claimed that FRB rates are higher ($\sim$10$^4$ sky$^{-1}$day$^{-1}$) than BNS merger rate \citep{2013Sci...341...53T, 2018ApJ...858...89C, 2019NatAs...3..928R, 2019JHEAp..23....1D, 2020MNRAS.494..665L}, but FRB rate estimates are highly uncertain and significantly different rates have been reported by different authors.  
In fact, several other studies report lower estimates ($\gtrsim$10$^3$ sky$^{-1}$day$^{-1}$) that is closer to the BNS merger rate \citep{2013MNRAS.436..371H, 2015MNRAS.447.2852K, 2017AJ....154..117L, 2016MNRAS.455.2207R, 2016MNRAS.460.3370C, 2017AJ....154..117L, 2018MNRAS.475.1427B, 2019A&ARv..27....4P, 2019ApJ...883...40L, 2020MNRAS.497..352A}. 
The observed FRB rates should include repeating population, and the rate of the one-off population, if exists, should be smaller.  
A study of FRB luminosity function found that FRB rate above 10$^{43}$ erg/s is compatible with the BNS merger rate, and interestingly, most of apparently non-repeating FRBs are brighter than 10$^{43}$ erg/s \citep{2020MNRAS.494..665L}.
}

\R{ The beaming factor of FRB radio emission is important for the rate comparison. 
If a significant fraction of observed FRBs are one-off events produced by BNS mergers, FRB emission cannot be strongly beamed from the rate comparison. Instead, non-repeating FRBs by BNS mergers may constitute only a small fraction of all FRBs.  
Assuming isotropic radio emission of FRBs is not unreasonable, because the beaming fraction of pulsar radio emission in general lies between 0.4 to unity \citep[][]{1993ApJ...408..637F, 1998ApJ...501..270K, 2005ApJ...625..796H, 2008LRR....11....8L, 2010ApJ...715..230O, 2010ApJ...716L..85R}, though some FRB theoretical models suggest a stronger beaming \citep[e.g.][]{2017MNRAS.467L..96K, 2020MNRAS.tmp.2233C, 2020arXiv200506736L}.  
}

The possible relation between FRBs and BNS mergers leads to potential association with GWs as well as multi-wavelength electromagnetic counterparts of the latter.  Such a hypothesis has been tested by deep targeted search for \R{GWs in coincidence with FRBs \citep{2016PhRvD..93l2008A}}, FRBs from GRB remnants \citep{2019ApJ...887..252M} and optical follow-up of FRBs \citep{2014PASJ...66L...9N, 2018PASJ...70L...7N, 2018PASJ...70..103T, 2020A&A...639A.119M, 2020ApJ...896L...2A}, but no significant candidate has been identified.  
Identification of any electromagnetic counterpart to an FRB will not only help pinpoint its host galaxy, but also provide crucial information on the progenitor system and ambient environment.

Throughout this paper we consider BNS mergers as the progenitor model for (both repeating and non-repeating) FRBs, and focus on the radio detectability of afterglows powered by the relativistic jet and the sub-relativistic ejecta in order to test the BNS merger model of FRBs.

% 2018ApJ...858...89C (compact merger vs FRB rate)

For calculation of radio afterglows from BNS mergers, we use the model developed by our previous work \citep{2019MNRAS.485.2155L}. 
A unique point of this model is treating the efficiency of electron acceleration and the minimum electron energy as independent parameters, while other previous models made an oversimplified assumption that all electrons in the shock is accelerated as nonthermal particles.
As a result of one more degree of freedom, we found a quantitatively different fit to the observed light curve of GW170817. 

The paper is organized as follows: in Section \ref{sec:model}, we describe the models of outflow dynamics and afterglow emission. In Section \ref{sec:result}, we present our assumptions for model parameter distributions, and quantify the detection probability as function of given source redshift and detector sensitivity. In Section \ref{sec:dis}, we compare the result to FRBs detected with radio limits and discuss implications for BNS progenitor models. The conclusion is given in the Section \ref{sec:con}. 
We adopted $H_0=67.4$ km/Mpc/s, $\Omega_{m}=0.315$, $\Omega_{\Lambda}= 0.685$ from \cite{2018arXiv180706209P}.

\begin{figure}
\begin{center}
    \includegraphics[width=1\columnwidth]{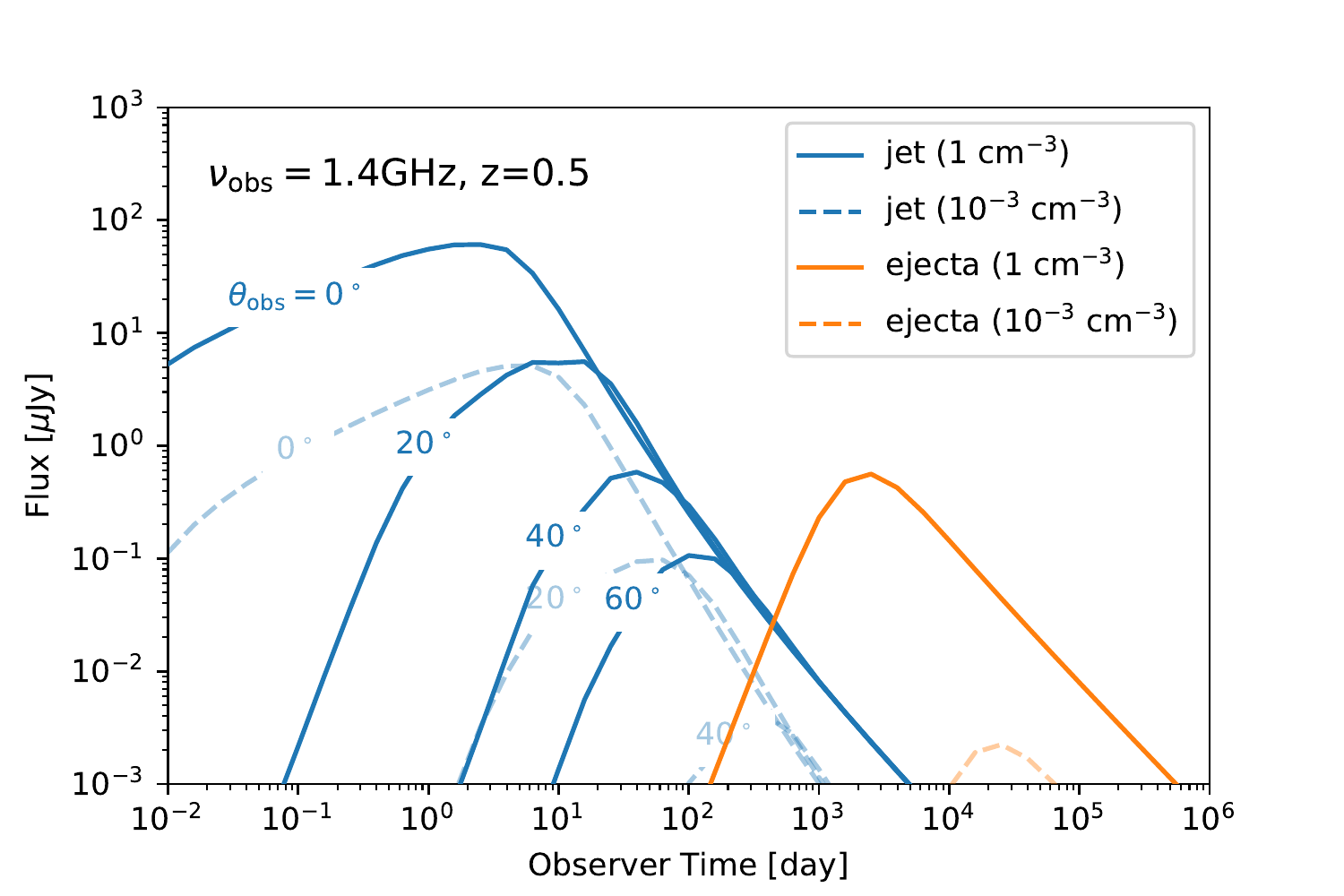}
    \caption{Radio (1.4 GHz) afterglow light curves of structured jet viewed in various angles (blue) and isotropic ejecta (orange) from a source at $z=0.5$. The solid lines account for the typical ambient density in a star forming galaxy (1\pcm) while dotted lines for an old elliptical galaxy (10$^{-3}$\pcm). 
    }
    \label{fig:lc}
\end{center}
\end{figure}

\section{Model} \label{sec:model}

We apply the same model in \cite{2019MNRAS.485.2155L} for both outflow dynamics and afterglow emission, as described below. Two outflow components are considered to power the afterglow emission: an axisymmetric jet with Gaussian angular profile of energy/Lorentz factor, and (quasi-)isotropic ejecta with power-law radial velocity stratification. The relevant model parameters include isotropic energy ($E_{k,\rm iso}$), initial Lorentz factor ($\Gamma_0$), ambient particle number density of hydrogen ($n$), angular width of the Gaussian jet ($\theta_j$), and power-law index of ejecta stratification ($\alpha$). Here $E_{k,\rm iso}$ and $\Gamma_0$ refer to the isotropic-equivalent energy and Lorentz factor of the jet head, as well as the total energy and common Lorentz factor of isotropic ejecta, respectively. Then the velocity $\beta(R)$ of shock into any direction can be solved as a function of its radius $R$. 

For the electron distribution developed in shock acceleration, we approximate the solution by broken power law \citep[see also][]{1998ApJ...497L..17S}. The relevant parameters include the fractions of internal energy transferred to electrons and magnetic field ($\epsilon_e$, $\epsilon_B$), power-law index of electron injection ($p$) and number fraction of injected electrons in the downstream ($f$). Using the electron distribution as input, the comoving synchrotron energy spectrum $P(\nu, R)$ is calculated as function of comoving photon frequency $\nu$ and shock radius $R$ using the public Python package {\tt naima} \citep{2015ICRC...34..922Z}. The effect of synchrotron self absorption is ignored since it does not affect our result significantly. 

Finally, the energy flux received by a distant observer is obtained by integrating the Doppler boost of $P(\nu, R)/(4\pi D_L^2)$ over the photon equal time arrival surface, which can be solved from $\beta(R)$ in all directions \citep[see also][]{1999ApJ...513..679G}, given the following observer parameters: observed photon frequency $\nu_{\rm obs}$, observed time $t_{\rm obs}$, viewing angle $\theta_{\rm obs}$, and luminosity distance $D_L$. 

Radio (1.4 GHz) afterglow light curves from a source located at $z=0.5$, the median redshift of detected FRBs \citep[e.g.][]{2016PASA...33...45P}, are shown in Fig.\ref{fig:lc} for illustration. The parameters for jet/ejecta components are chosen as follows: $E_{k,\rm iso} = 10^{51}$ erg, $\Gamma_0=100$, $\theta_j = 6^\circ$, and viewing angles ranging from $0$ to $0.5$ for jet; $E_{k,\rm iso} = 10^{51}$ erg, $\Gamma_0=1.05$, $\alpha\to\infty$ (i.e. single velocity structure) for ejecta. The common parameters are $\epsilon_e=0.1$, $\epsilon_B=0.01$, $p=2.2$, $f=1$, and the typical ambient densities in a star forming galaxy ($1$\pcm) and an old elliptical galaxy ($10^{-3}$\pcm), respectively \citep{2018ApJ...865...27W}. 

\section{Result} \label{sec:result}

\R{We assume as follows the distributions for model parameter set ($E_{k, \rm iso}$, $\Gamma_0$, $n$, $\epsilon_e$, $\epsilon_B$, $p$, $f$, $\theta_j$, $\theta_{\rm obs}$, $\alpha$).}
We fix the Lorentz factor $\Gamma_0$ to fiducial values: $100$ for jet and $1.05$ for ejecta, and adopt the Gaussian fits (Table.\ref{tab:F15}) to the cumulative distributions of ($E_{k,\rm iso}$, $n$, $\theta_j$, $p$) inferred from short GRB afterglow observations \citep[][F15]{2015ApJ...815..102F}, with fixed values of $\epsilon_e=0.1$, $\epsilon_B=0.01$, $f=1$ and a lower bound of $n_{\rm min}=10^{-4}$\pcm for the $n$ distribution imposed by F15. 
The distribution of $\theta_j$ is estimated using only GRBs showing jet breaks in their afterglows. 
As no observational constraint is available for the scatter of ejecta parameter, we fix $E_{k,\rm iso}=10^{51}$ erg for ejecta as reference to the kinetic energy of kilonova ejecta seen in GW170817 \citep{2017ApJ...851L..21V} and $\alpha\to\infty$ (i.e. single velocity structure) for simplicity, also considering the observational constraint $\alpha>6$ suggested from late time radio monitoring of GW170817 \citep{2019ApJ...886L..17H}.

\R{For the jet viewing angle $\theta_{\rm obs}$, a general assumption of the probability density distribution (PDF) would be the uniform spherical distribution ($dP/d\theta_{\rm obs} = \sin\theta_{\rm obs}$, hereafter uniform PDF), i.e. observers are uniformly distributed on a sphere, which accounts for the detection probability to follow up an FRB (assuming FRB is not strongly beamed) or in an untargeted search. 
For completeness, we also consider cases such as targeted follow-up of short GRBs and GWs from BNS mergers, for which the detections are not drawn from a uniform PDF. 
For example, short GRBs are highly beamed and therefore the detection are heavily biased towards on-axis bursts ($\theta_{\rm obs}\lesssim \theta_j$); the gravitational wave detectors also have a greater detector range for small observing angles \citep[][hereafter Schutz PDF]{2011CQGra..28l5023S}.
Therefore, in this work we assume 3 PDFs in total to reflect different searches: uniform PDF, Schutz PDF and on-axis (fixing $\theta_{\rm obs}=0$), to reflect these different searching scenarios respectively. } 

\begin{figure*}
\begin{center}
\begin{multicols}{2}
    \includegraphics[width=1\columnwidth]{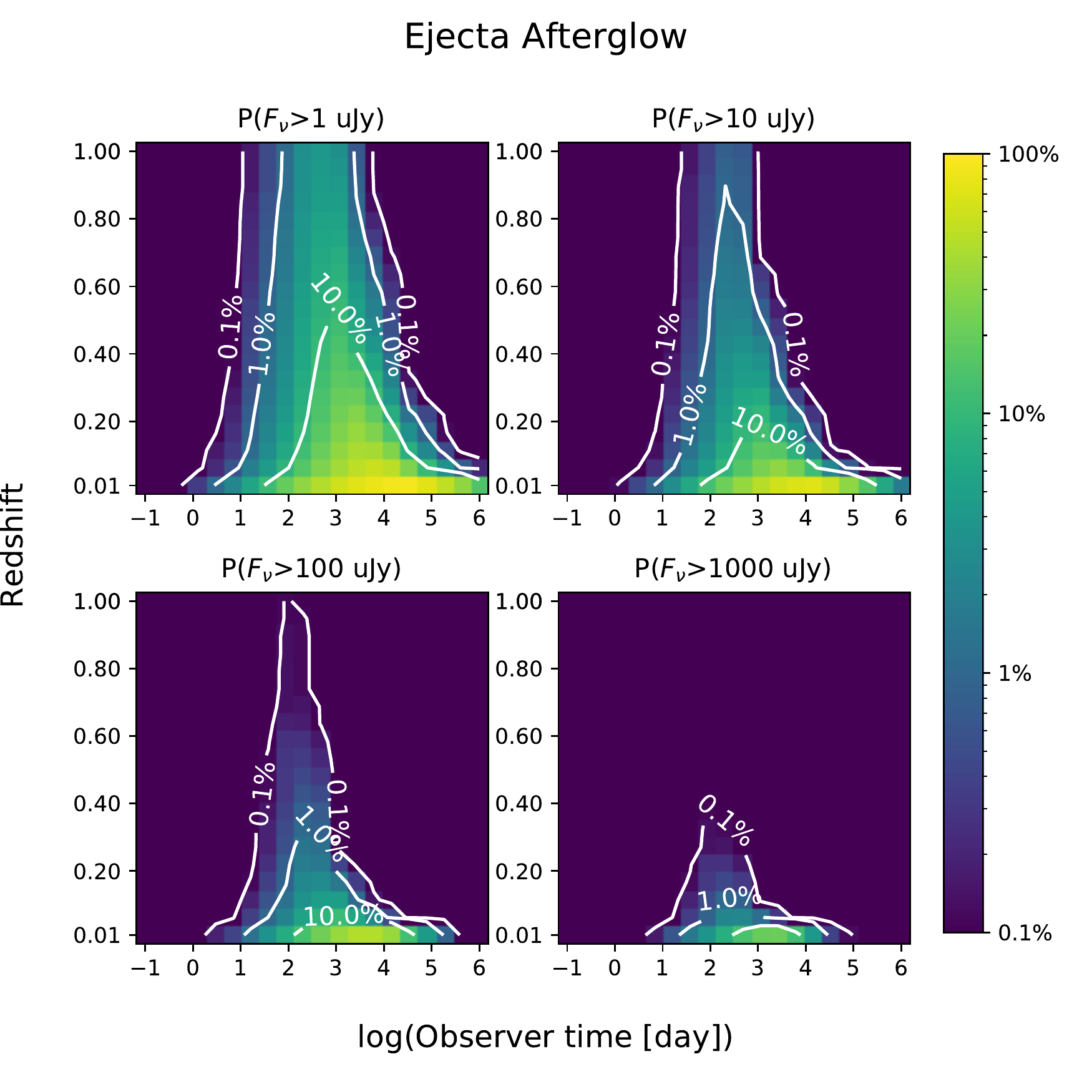}%\par
    \includegraphics[width=1\columnwidth]{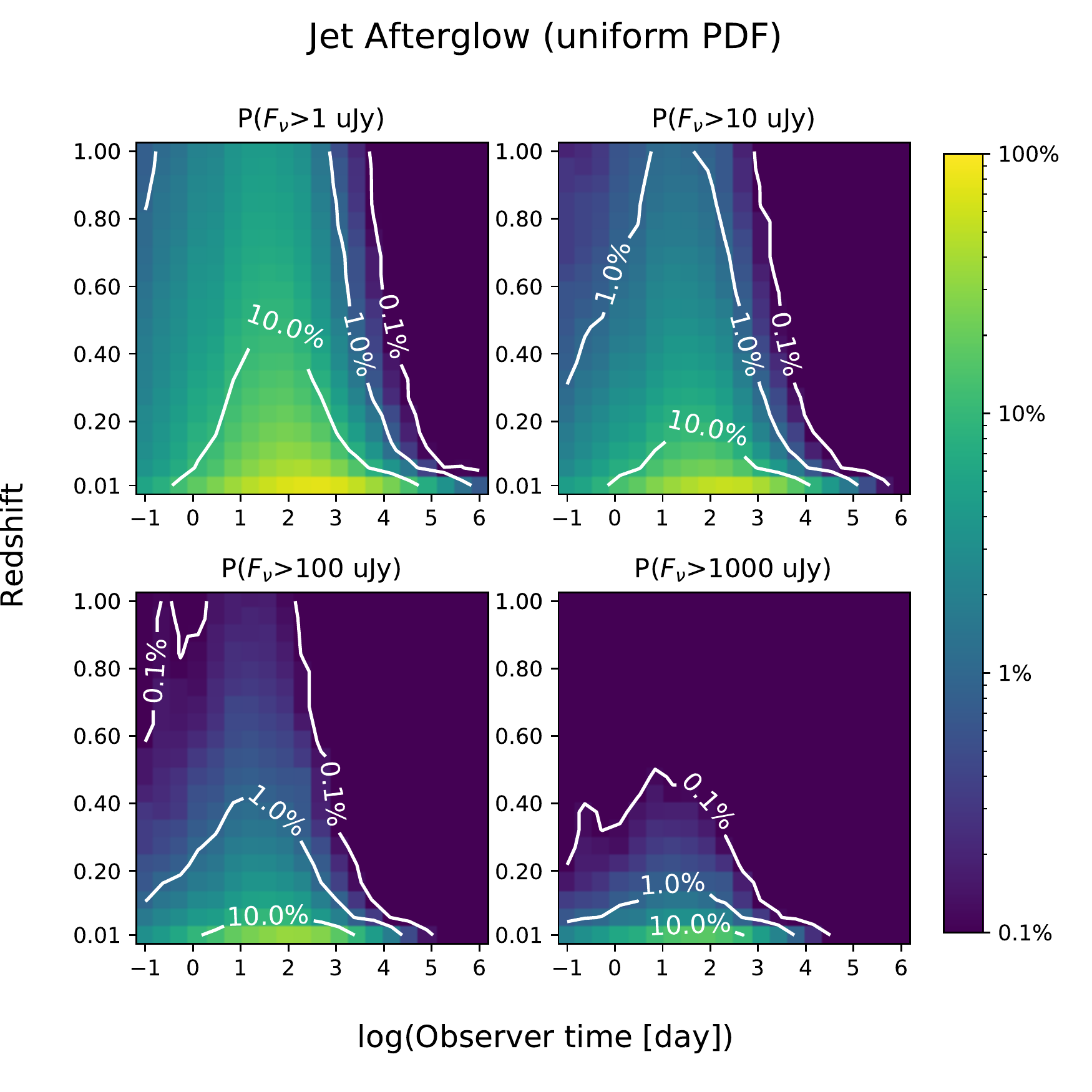}%\par
\end{multicols}
\begin{multicols}{2}
    \includegraphics[width=1\columnwidth]{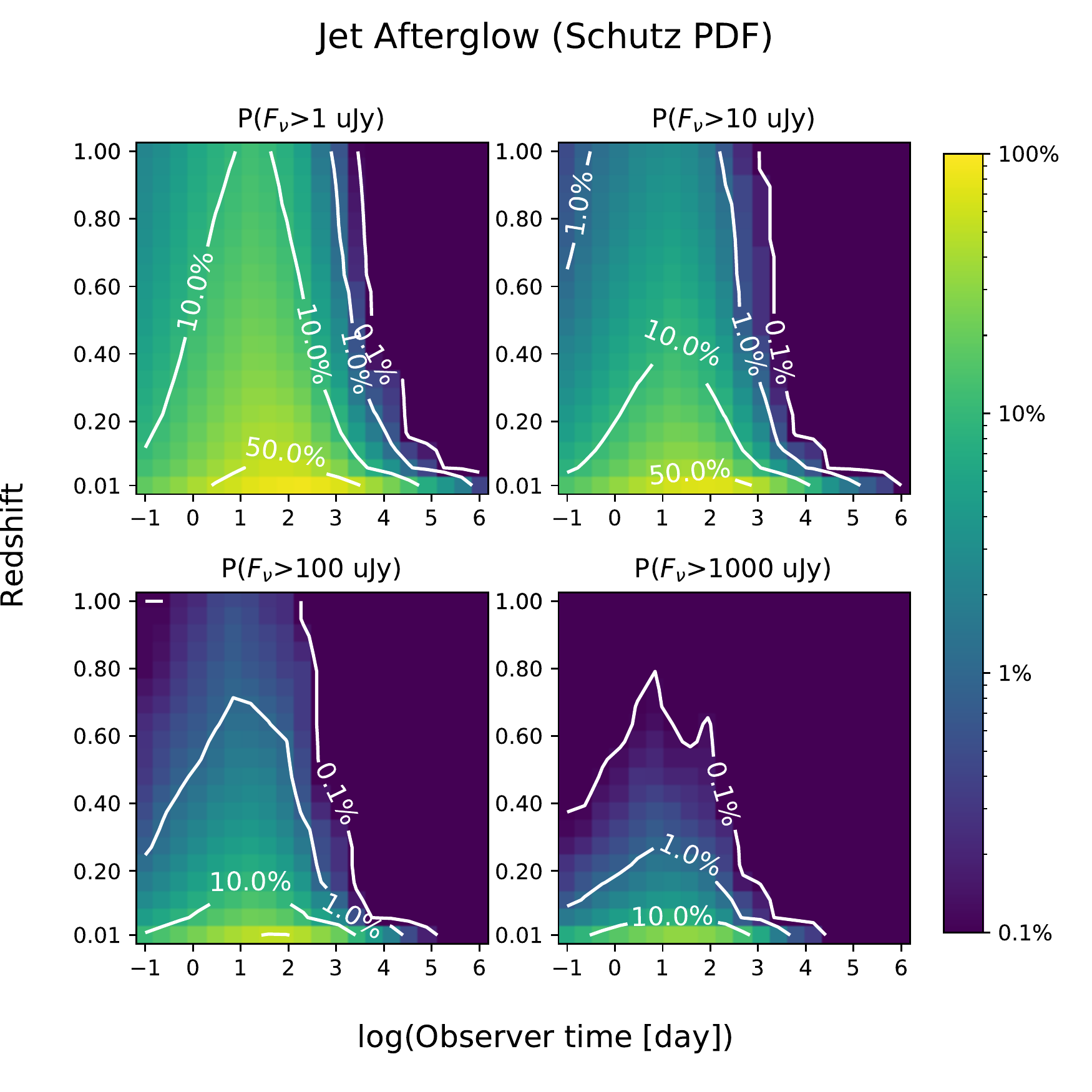}%\par
    \includegraphics[width=1\columnwidth]{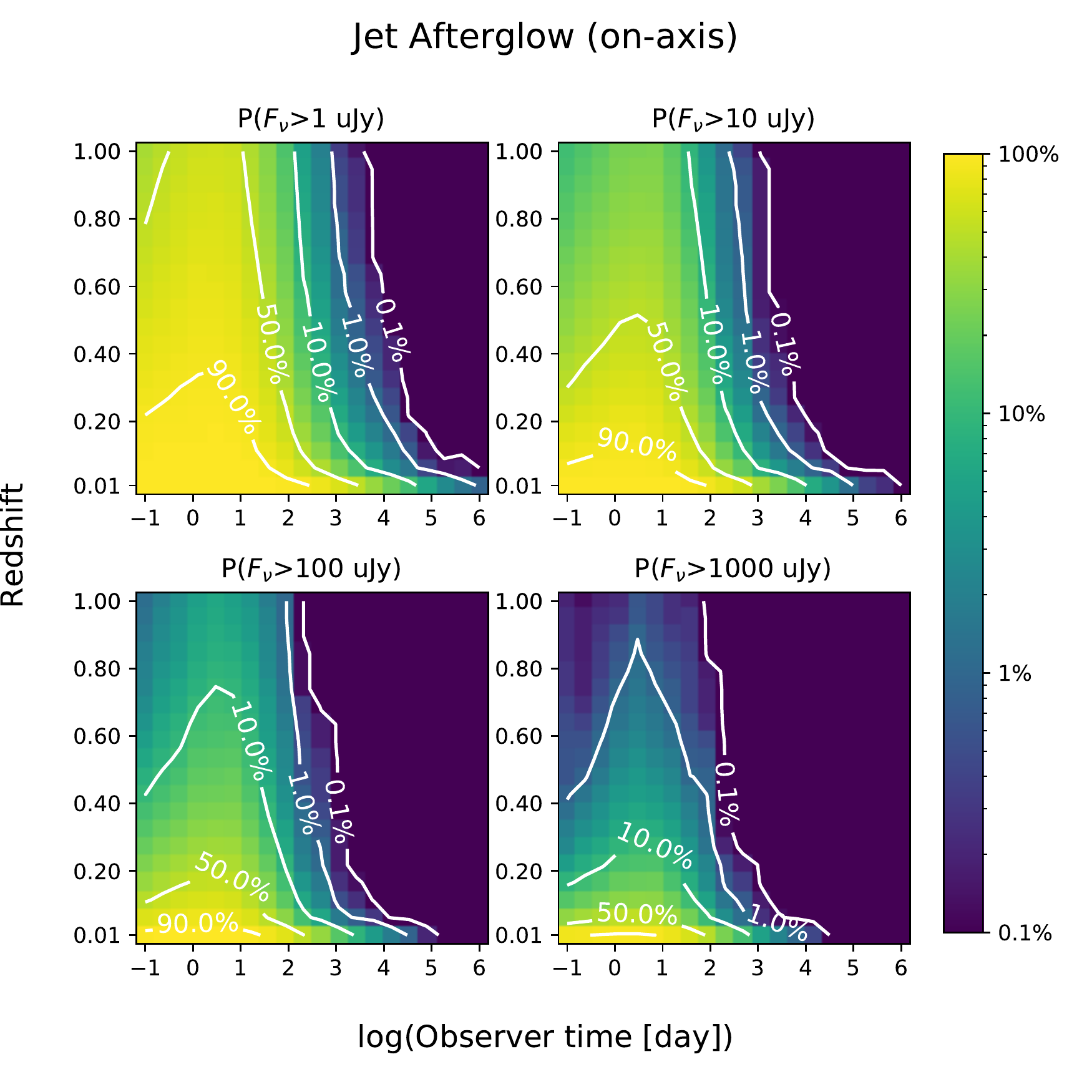}%\par
\end{multicols}
\caption{Detection probabilities of (1.4 GHz) radio afterglow powered by the isotropic ejecta (upper left) and jet assuming uniformly-distributed viewing angle (upper right) at a detector sensitivity of 1, 10, 100, and 1000$\mu$Jy, as a function of source redshift and observation time. Lower panels are the same by the jet component, but assuming Schutz distribution (lower left) and heavily biased on axis (lower right), respectively. Contours of detection probabilities are shown. Color scale below 0.1\% is not shown due to the insufficient sensitivity of the calculation.}
\label{fig:prob}
\end{center}
\end{figure*}

Producing 10$^{3-4}$ model parameter sets \R{(as the probability converges at a set number $\sim $10$^3$)} randomly from the distributions above, we calculated radio light curves for each model set, and quantify the detection probability of radio afterglow as the observable fraction in these sets as a function of a given detection threshold $F_{\rm lim}$, a source redshift $z$, and the observation time $t_{\rm obs}$. 
The results for jet (uniform, Schutz, on-axis cases of $\theta_{\rm obs}$ distributions) and ejecta are shown in Fig.\ref{fig:prob}, where the observed frequency is fixed at 1.4 GHz. 
\R{The flux density has a relatively weak scaling with the observed frequency as $F_\nu \sim \nu_{\rm obs}^{-(p-1)/2}$ (e.g. $F_\nu \sim \nu_{\rm obs}^{-0.6}$ in GW170817) in the optically-thin regime, and hence the results can be generalized and compared to other surveys, e.g. the VLA Sky Survey at 3 GHz and ASKAP transients surveys at 900 MHz.}

\begin{figure*}
\begin{center}
\begin{multicols}{2}
    \includegraphics[width=1\columnwidth]{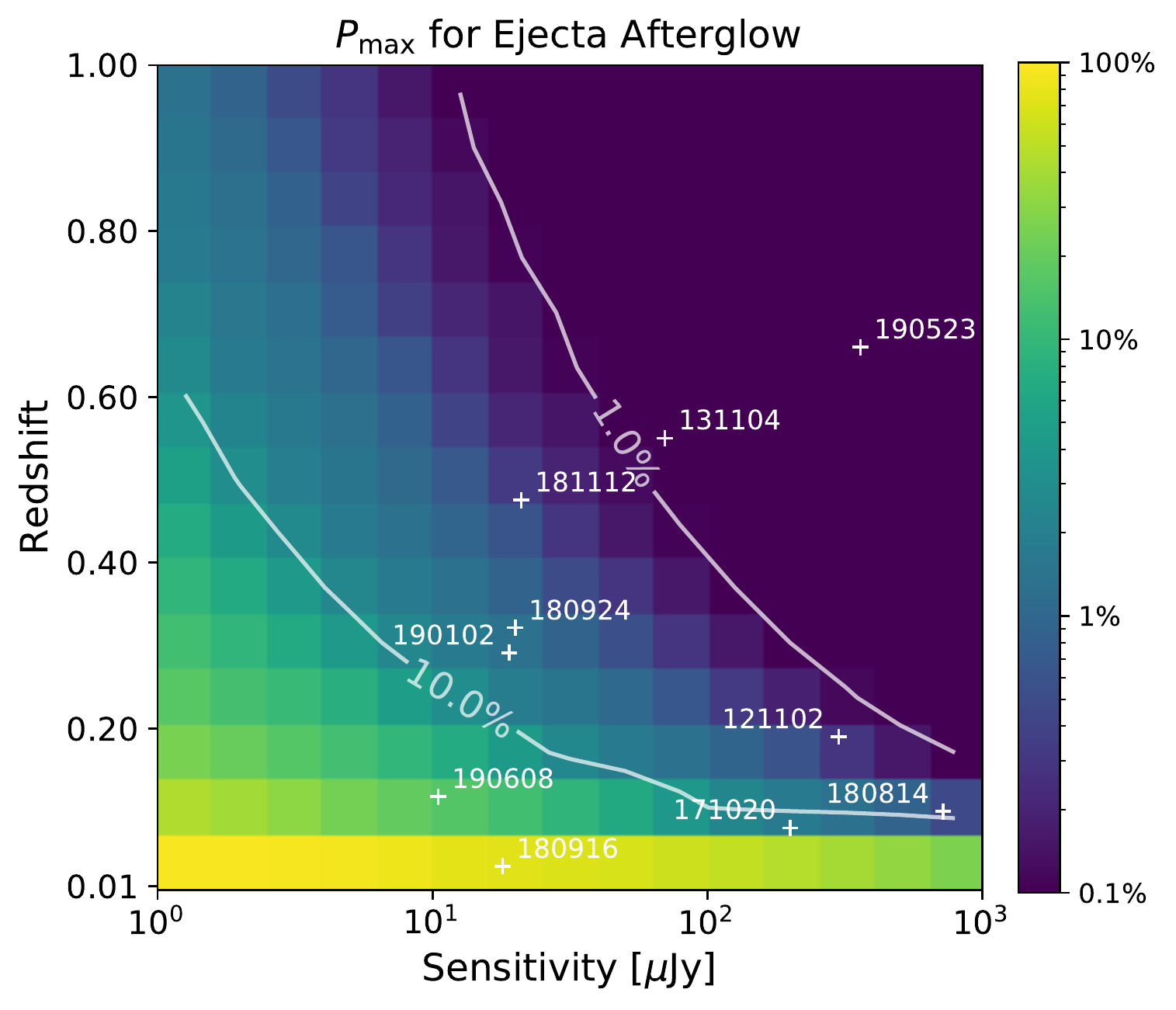}%\par
    \includegraphics[width=1\columnwidth]{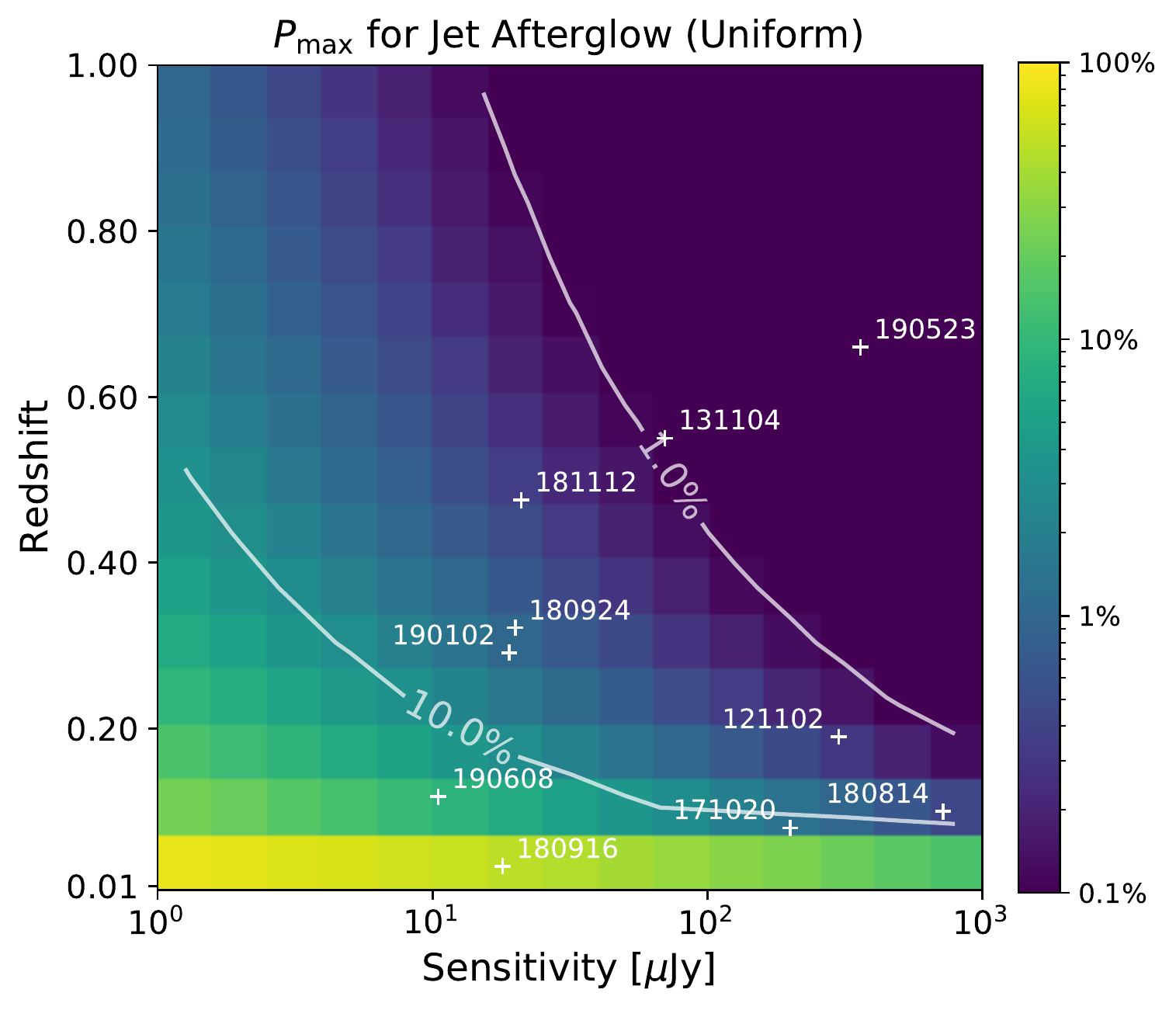}%\par
\end{multicols}
\begin{multicols}{2}
    \includegraphics[width=1\columnwidth]{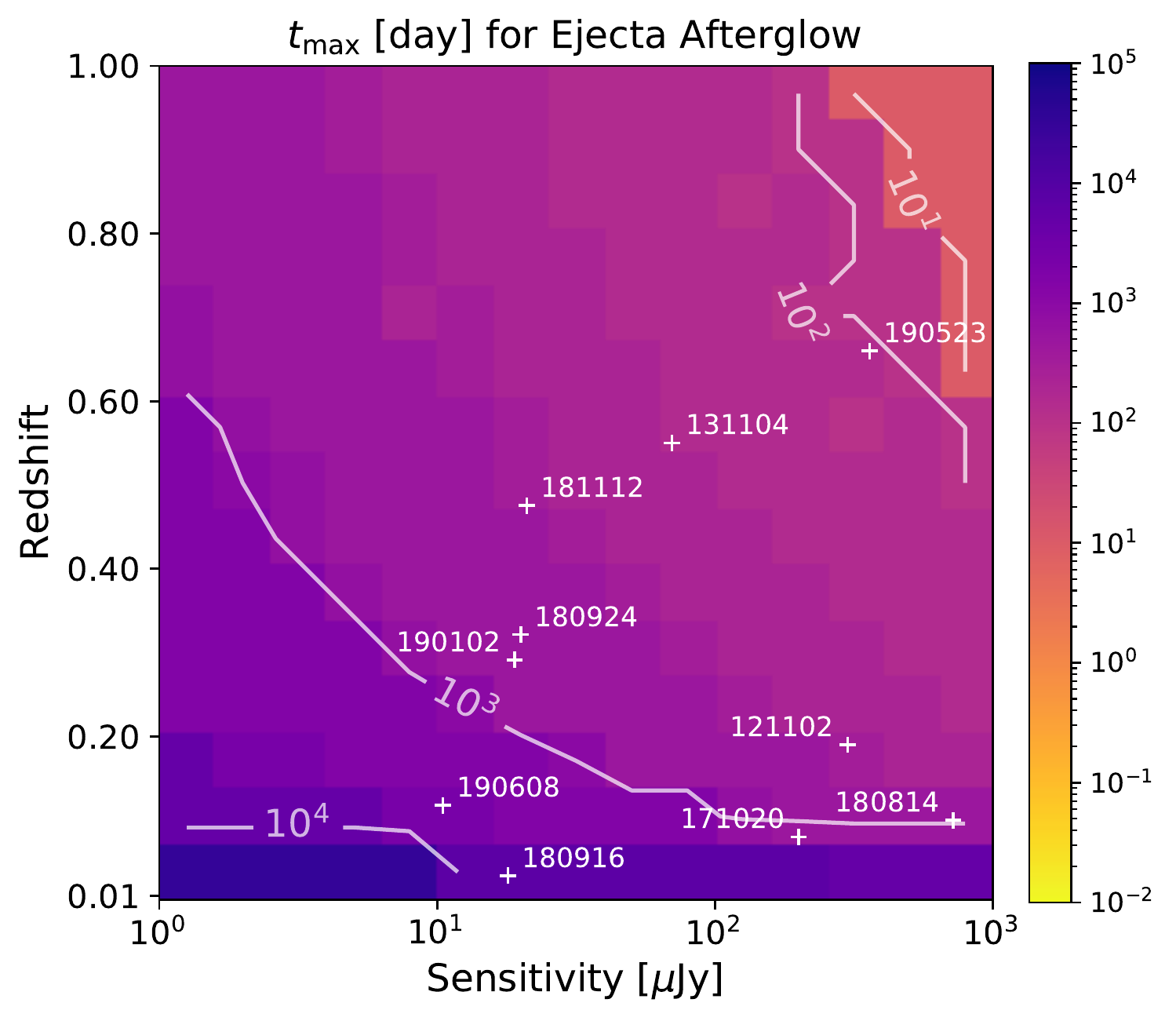}%\par
    \includegraphics[width=1\columnwidth]{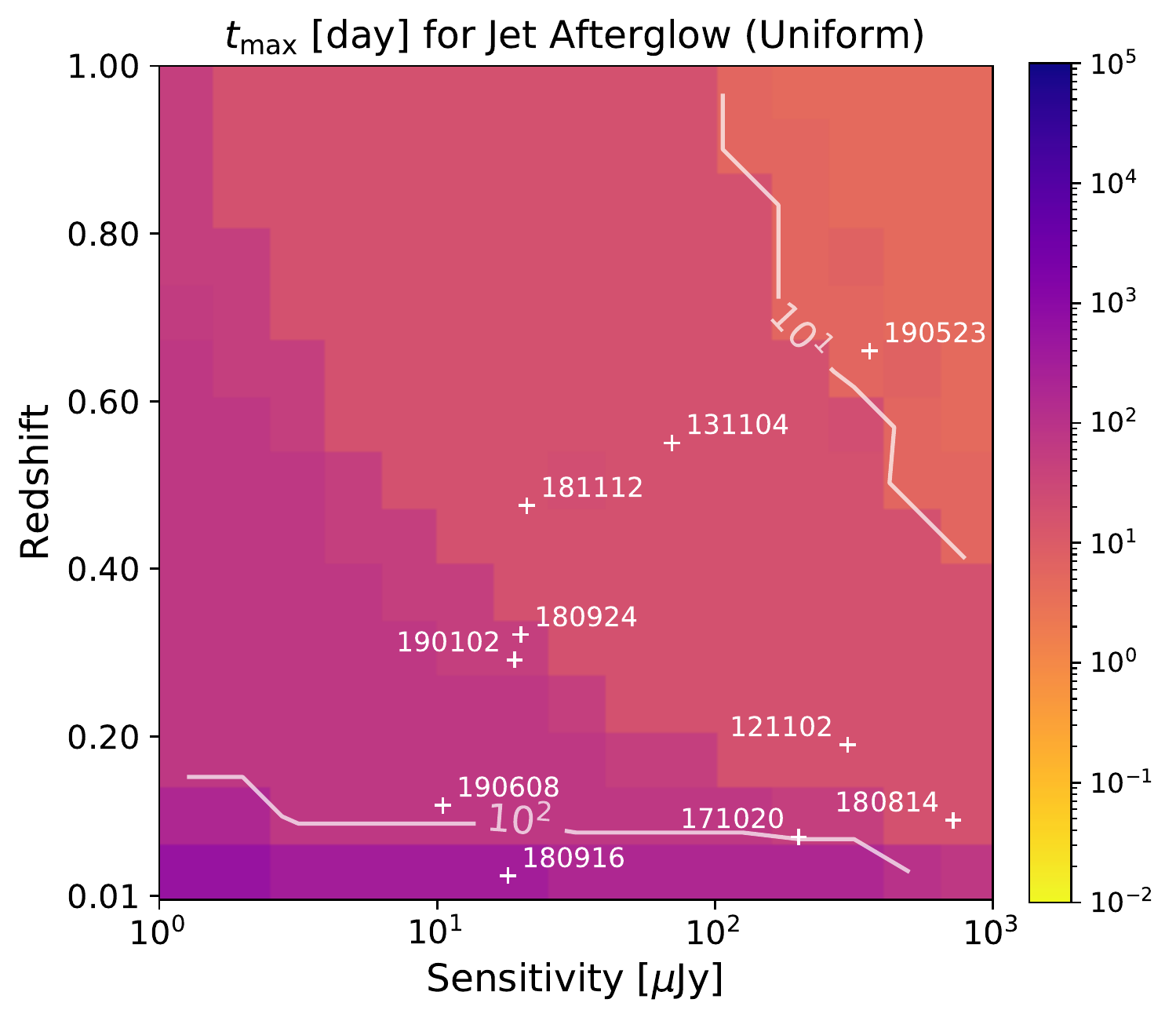}%\par
\end{multicols}
\caption{The maximized detection probability of radio afterglow of all time (upper) and the corresponding observation time (lower) as a function of redshift and detector sensitivity. 
FRBs with reported upper limits on a persistent radio emission listed in Table.\ref{tab:frb} are shown by plus marks.
FRBs are assumed not to be strongly beamed towards any favored direction from the jet axis, and hence are only compared to the cases where viewing angle is uniformly distributed.
}
\label{fig:max}
\end{center}
\end{figure*}

\R{To compare with observational radio limits on FRB, we assume that the FRB emission is not strongly beamed, or at least not favored towards a particular direction from the jet axis so that $\theta_{\rm obs}$ is randomly drawn from uniform PDF, and thus the radio constraints on FRB afterglow should be compared with those results for jet (uniform PDF) and isotropic ejecta afterglow. }
\R{To make comparison, we calculate the maximized detection probability of all time $P_{\rm max}$ and the corresponding time $t_{\rm max}$ for jet (uniform PDF) and ejecta afterglow, showing in Fig.\ref{fig:max} as a function of redshift and detector sensitivity. }
FRBs with reported upper limits on radio afterglows \citep{2016ApJ...833..177S, 2017ApJ...837L..22S, 2018ApJ...867L..10M, 2019Sci...365..565B, 2019Natur.572..352R, 2019Natur.566..235C, 2019Sci...366..231P, 2020Natur.577..190M, 2020ApJ...895L..37B} are marked and also listed in Table.\ref{tab:frb}.

\begin{table}
\begin{center}
\begin{tabular}{|c|c|c|}
\hline
Parameters & Median & Standard Deviation \\ \hline
$\log(E_{\rm k,iso}\,[{\rm erg}])$ & 51.2   & 0.83    \\ \hline
$\log(n\,[{\rm cm^{-3}}])$ & -1.57  & 1.63    \\ \hline
$\theta_j\,[{\rm deg}]$     & 5.8   & 1.2    \\ \hline
$p$          & 2.39   & 0.23    \\ \hline
\end{tabular}  
\end{center}
\caption{Median and 1$\sigma$ scatter of Gaussian fit to the parameter cumulative distributions by short GRB afterglow observation \citep{2015ApJ...815..102F}. }
\label{tab:F15}
\end{table}

\begin{table*}
\def\arraystretch{1.5}%
\caption{Radio afterglow detection probabilities for FRBs with reported upper limits $F_{\rm lim}$ on a persistent radio emission.}
\begin{center}
\begin{threeparttable}
\begin{tabular}{|c|c|c|c|c|c|c|c|p{0.18\textwidth}|}
\toprule
Source & $z\tnote{a}$ & detector & $\nu_{\rm obs}$ [GHz] & $F_{\rm lim}$ [$\mu$Jy]\tnote{b} & $t_{\rm lim}$ [day]\tnote{b} & $P_{\rm jet}$ ($P_{\rm ej}$)\tnote{c} & $t_{\rm jet}$ ($t_{\rm ej}$) [day]\tnote{c} & Reference \\ \hline
% FRB 121102 was observed during +1118d--1140d from date 121102 (on point-source continuum emission)
FRB 121102\tnote{*} & 0.19 & VLA & 1.6 & 300 (5$\sigma$) & 1117 & 2\% (2\%) & 20 (280) & \cite{2016ApJ...833..177S} \\ \hline
FRB 180814\tnote{*} & 0.1 (max) & VLA & 3 & 720 (5$\sigma$) & - & 3\% (3\%) & 20 (380) & \cite{2019Natur.566..235C} \\ \hline
\multirow{2}{*}{FRB 180916\tnote{*}} & \multirow{2}{*}{0.0337} & EVN & 1.7 & 30 (3$\sigma$) & 275 & 41\% (55\%) & 140 (5000) & \multirow{2}{*}{\cite{2020Natur.577..190M}} \\ \cline{3-8}
& & VLA & 1.6 & 18 (3$\sigma$) & - & 45\% (59\%) & 215 (5000) \\ \hline
\hline
FRB 131104 & 0.55 (max) & ATCA & 5.5 & 70 (5$\sigma$) & 3--900 & 1\% (0.7\%) & 20 (190) & \cite{2017ApJ...837L..22S} \\ \hline
FRB 171020 & 0.08 (max) & ATCA & 2.1 & 200 (5$\sigma$) & 218 & 6\% (7\%) & 50 (520) & \cite{2018ApJ...867L..10M} \\ \hline
\multirow{2}{*}{FRB 180924} & \multirow{2}{*}{0.3214} & ATCA & 6.5 & 20 (3$\sigma$) & 1--10 & 4\% (5\%) & 30 (460) & \multirow{2}{*}{\cite{2019Sci...365..565B}} \\ \cline{3-8} 
 & & ASKAP & 1.3 & 450 (3$\sigma$) & 2 & 0.6\% (0.5\%) & 20 (170) \\ \hline
FRB 190523 & 0.66 & VLA & 3 & 360 (3$\sigma$) & - & 0.2\% (0.06\%) & 6 (140) & \cite{2019Natur.572..352R} \\ \hline
FRB 181112 & 0.4755 & ATCA & 6.5 & 21 (3$\sigma$) & 5 & 2\% (2\%) & 20 (350) & \cite{2019Sci...366..231P} \\ \hline
FRB 190102 & 0.2913 & ATCA & 6.5 & 19 (3$\sigma$) & 69 & 4.7\% (5.7\%) & 46 (460) & \cite{2020ApJ...895L..37B} \\ \hline
FRB 190608 & 0.11778 & ATCA & 6.5 & 10.5 (3$\sigma$) & 74 & 15\% (23\%) & 68 (1750) & \cite{2020ApJ...895L..37B} \\
\bottomrule
\end{tabular}
\begin{tablenotes}\footnotesize
\item[*] Repeating FRB sources
\item[a] Redshifts inferred from localized host galaxies, except those of FRB 180814, 131104 and 171020 which were inferred as the maximum values from their dispersion measures (see corresponding references).
\item[b] Upper limits on the possible persistent radio emission and corresponding observation times after FRB detection (or after detection of the first burst in the case of repeater). VLA limits for FRB 180814, 180916 and 190523 were obtained based on the non-detection in the VLA Sky Survey performed in 2017 ({\url{https://science.nrao.edu/vlass}}), i.e., prior to the FRB detection.
\item[c] The maximized detection probability of all time at the sensitivity of $F_{\rm lim}$ and the corresponding peak observation time, for the jet and isotropic components (the latter in parentheses).

\end{tablenotes}
\end{threeparttable}
\end{center}
\label{tab:frb}
\end{table*}
% FRB 190523: VLASS obs. on 2017-11-24 (T28t07)
% FRB 180916: VLASS obs. on 2017-11-05 (T27t02)
% FRB 180814: VLASS obs. on 2017-09-19 (T29t02)

\section{Discussion} \label{sec:dis}

% designed 1$\sigma$ rms (2$\mu$Jy) of SKA-1 at 1.3 GHz \citep{2019arXiv191212699B}

\subsection{Detection prospects of radio afterglows}
\R{For both jet (uniform PDF) and isotropic ejecta sources beyond $z=0.2$, we found that the detection probability of radio afterglow is generally less than 10\%, even at the best sensitivity of current radio telescopes ($\sim 10\mu$Jy). 
The probability peaks typically at $\sim$10--100 days for a jet afterglow and $\sim$1--10 year for an ejecta afterglow. 
In the case of jet (Schutz PDF) relevant to GW follow-up, the 10\% detection contour is extended to $z=0.4$. and to include all sources within $z<1$ in the case of jet (on-axis) as the detection becomes more biased to axis. We note that the predicted high detection probability in the jet (on-axis) case should be considered only as an upper bound for any targeted search, since even GRBs can still be observed off-axis as faint GRBs \citep[e.g.][]{2018PTEP.2018d3E02I, 2018NatCo...9.4089T}. }

Note that although the jet afterglow seems brighter than the ejecta afterglow in the light curves (Fig.\ref{fig:lc}), viewing angles in the presented range only constitute 50\% of the whole population if assuming uniform PDF. At the median value ($\langle\theta_{\rm obs}\rangle=60^\circ$), most jets are viewed heavily off-axis and produce comparable afterglow flux level to ejecta, which results in the similar detection probability (but different peak time) seen in the upper panels of Fig.\ref{fig:prob}. 

\R{These results can be compared with radio afterglow search of short GRBs: (i) 2 early (1--10 days) radio afterglow detections out of 35 short GRBs \citep{2012ApJ...746..156C}, which implies a jet afterglow detection probability of $\sim 6\%$; (ii) no late time detection of radio afterglows from 16 short GRBs (2005--2015) with $15$--$40\mu$Jy (3$\sigma$) upper limit \citep{2019ApJ...887..206K}, and from another 9 samples (2.4--13.9 yr) with $6$--$20\mu$Jy (3$\sigma$) upper limit \citep{2020arXiv200607434S}, which in total implies an ejecta afterglow detection probability of $\lesssim 4\%$. And hence the observational detection rates are consistent with our result within an order of magnitude.
}

% 2018MNRAS.474.5340S: similar work on jet afterglow (assuming Schutz PDF for GW), but what they do is searching parameter space of detection

\subsection{Consistency with radio limits on FRBs}

We discuss whether or not the available limits on persistent radio emission imposed on some FRBs by follow-up observations is consistent with our calculated detection probability. 
From Fig.\ref{fig:max} and Table \ref{tab:frb}, we found that the detection probabilities for most FRBs are typically in the range of 1--10\%. 
Considering that the total number of the available sample is less than 10 FRBs, the non-detection of any associated radio afterglow does not strongly constrain the BNS progenitor model.

\R{FRB 190608} and the repeater 180916.J0158+65 are the only 2 exceptions with a promising detection prospect ($23\%$ and $60\%$) for the isotropic component that peaks around $\sim$1--10 years after the merger, as a natural consequence of their close distances. 
Yet the probability does not go up to 100\% mainly because of the low ambient densities that may appear in the adopted parameter distribution.
We further note that the repeating FRB activity would start 1--10 years after a BNS merger \citep{2018PASJ...70...39Y}. 
Therefore there is a reasonable chance to detect a radio afterglow for FRB 180916.J0158+65 in the near future, though no detection does not give a strong constraint.

The ambient density distribution adopted in this work from F15 is widely spread from $10^{-3}$ to $1$ cm$^{-3}$. However, some FRBs listed in Table.\ref{tab:frb} are well localized to a type-specified host galaxy. 
Therefore, the detection probability is underestimated if the source is localized in high density environment like star-forming region (e.g. FRB 180916.J0158+65), and overestimated if localized in a more diffuse environment like an early-type galaxy or outskirt regions (e.g. FRB 180924, 190523).

Note that Fig.\ref{fig:max} shows a tendency of later peak time $t_p$ towards closer distance and higher sensitivity. 
The key to interpret this result is the scaling relation of afterglow peak flux and peak time: $F_p \propto E_{\rm k,iso} n^{(p+1)/4}$, $t_p \propto (E_{\rm k,iso}/n)^{1/3}$. 
If the scatter in $E_{\rm k,iso}$ is negligible, the scatter in ambient density $n$ leads to a negative correlation $F_p \propto t_p^{-(3p+3)/4}$, i.e. fainter event with later peak time. 
This negative correlation remains true as long as $E_{\rm k,iso}$ is less scattered compared with $n$, which is the case for short GRBs (Table \ref{tab:F15}), though the correlation is weaker due to the finite scatter of $E_{\rm k,iso}$. 
Consequently, a higher sensitivity or a closer source will allow much fainter detections, which eventually leads to the tendency of later peak time.

\subsection{Caveats on the jet opening angle}

The distribution of jet opening angle $\theta_j$ assumed in this work significantly affects the detection probability as it scales with $\theta_j^2$. 
We note one important caveat that while we assumed the same parameter distributions for short GRBs and FRB afterglows in the BNS merger progenitor model, the parameter space in which an FRB is produced may not coincide with that for successful GRB jet
formation. 
One such scenario is that FRBs may appear in a BNS merger with a ``choked'' or failed jet in the parameter space of smaller $(E_{\rm k,iso})_{\rm jet}$ and larger $\theta_j$, e.g. $(E_{\rm k,iso})_{\rm jet} < 0.05(E_{\rm k,iso})_{\rm ej}\theta_j^2$ \citep{2018ApJ...866....3D}, which may produce a bias against afterglow detections at larger viewing angle ($\lesssim$ 23$^\circ$).
However, the resulting detection probability does not necessarily increase because the expected flux is fainter at a larger viewing angle.

Another caveat is that we have used the $\theta_j$ distribution of F15 based on 4 short GRBs with jet break measurements, which is narrowly peaked at 6$^\circ$ with a $1\sigma$ scatter of 1$^\circ$. 
F15 also provided another distribution by including 7 more samples with lower bounds on $\theta_j$ and assuming equal weighting over all 11 events, which extends the cumulative distribution almost linearly to an ad hoc maximum angle (30$^\circ$, 90$^\circ$). 
Considering the uncertainty in the treatment of these additional events, and the fact that the value of narrow $\theta_j$ distribution is well consistent with the jet width ($\theta_j\sim 5^\circ$) seen in GW170817 \citep{2019MNRAS.489.1919T, 2019ApJ...886L..17H}, we only adopt the narrow distribution model in this work as a conservative estimate for the detection probability (since it scales with $\theta_j^2$).

The final remark is that $\theta_j$ in F15 are measured as the jet edge of a top-hat jet, different from the definition of jet width in our Gaussian jet model. 
For a same jet break time observed in a short GRB afterglow, the Gaussian jet model will measure a slightly smaller $\theta_j$ (roughly by a factor of 0.7). 
However, the resulting difference of detection probability is at most a factor of 2, and hence we ignored it.

\section{Conclusion} \label{sec:con}

In this work we examined the prospects of detecting radio afterglows from BNS mergers, and compares them to radio limits on FRB afterglows. 
We considered the two components of outflow: a relativistic jet and a mildly relativistic and isotropic ejecta. 
Detection probabilities of radio afterglow were calculated as a function of sensitivity, source redshift and an observation time (Fig.\ref{fig:prob}), assuming viewing angle distributions (uniform PDF, Schutz PDF, biased to on-axis detection only) to reflect different search scenarios and adopting the model parameter distributions inferred from short GRB observations.

For uniform spherical distribution, we found that the detection probability from sources within $z=0.2$--$1$ is between 1--10\% for the best sensitivity ($10\mu$Jy) of current radio telescope. The 10\% detection probability contour is extended to $z=0.4$ in the case of Schutz PDF relevant to GW follow-up, and to include all sources within $z<1$ for heavy detection bias to on-axis burst, though the latter could be an overestimate for GRB follow-up since close sources can be observed as off-axis faint GRBs. The expected flux peaks typically at $\sim$10--100 days for a jet afterglow and $\sim$1--10 year for an isotropic afterglow. We also found a tendency of later peak time towards closer source and higher sensitivity, which can be attributed to the contribution from low luminosity events whose afterglows peak at a later time (Fig.\ref{fig:max}). 

For individual FRBs with reported upper limits on a persistent radio flux, we listed their detection probabilities of BNS merger radio afterglow for the two components in Table.\ref{tab:frb}. 
The maximum detection probability of all time is found less than 10\% for most of these FRBs, and hence no detection does not give a strong constraint on the BNS merger scenario of FRBs considering the current small sample number ($\lesssim$10) of radio-constrained FRBs.
However, a future larger sample \R{or higher sensitivity search \citep[e.g. designed noise level $\sigma_{\rm rms} = 2\mu$Jy for SKA-1 at 1.4 GHz,][]{2019arXiv191212699B}} would give a meaningful constraint or lead to a detection of a radio afterglow. 

Nevertheless, we note that in particular for the FRB 190608 and repeating FRB 180916.J0158+65, we found 23\% and 60\% chance of detection for the isotropic component respectively, whose fluxes peak at about 10 years after the merger and remain detectable for a few decades, as a natural consequence of their close distance.  
The time scale of 10 yrs is also comparable to the lifetime of repeating FRBs formed by a BNS merger.  
Though the detection probability is not close to 100\% because of the distribution of model parameters, a long-term radio monitoring of this object is thus interesting.

\section*{Acknowledgements}
HL is supported by the Japanese Government (MEXT) Scholarship and the JSPS Research Fellowships for Young Scientists (KAKENHI Grant Number 20J12200). TT is supported by the JSPS/MEXT KAKENHI Grant Numbers 18K03692 and 17H06362.

\section*{Data availability}
The data underlying this article will be shared on reasonable request to the corresponding author.

\bibliographystyle{mnras}
\bibliography{refs}

\bsp
\label{lastpage}
\end{document}